\documentclass[twocolumn,aps,prl,showpacs,preprintnumbers,amsmath,amssymb]{revtex4}
\usepackage{graphicx}
\usepackage{dcolumn}
\usepackage{bm}
\begin{document}
\title{Air transparent soundproof window}
\author{Sang-Hoon  \surname{Kim}$^{a}$} \email{shkim@mmu.ac.kr}
\author{Seong-Hyun  \surname{Lee}$^{b}$} \email{sh.lee@kimm.re.kr}
\affiliation{
$^a$Division of Marine Engineering, Mokpo National Maritime University,
Mokpo 530-729, R. O. Korea
\\
$^b$Korea Institute of Machinery and Materials,
 Yuseong-Gu, Daejeon 305-343, R. O. Korea
}
\begin{abstract}
A soundproof window or wall which is transparent to airflow is presented.
The design is based on two wave theories of diffraction and acoustic metamaterials.
It consists of a three-dimensional array of strong diffraction-type resonators with many holes centered at each individual resonator.
The acoustic performance levels of two soundproof windows with air holes of
$20mm$ and $50mm$ diameters were measured.
Sound waves of 80dB in the frequency range of $400 - 5,000Hz$ were applied to the windows.
It was observed that the sound level was reduced by about $30 - 35$dB
in the above frequency range with the $20mm$ window and by
about $20 - 35$dB in the frequency range of $700 - 2,200Hz$
 with the $50mm$ window.
 It is an extraordinary acoustic anti-transmission.
 The geometric factors which produced the effective negative modulus were obtained.

\end{abstract}
  \maketitle

Electromagnetic waves are transverse waves and do not need any medium to travel.
On the other hand, sound waves are longitudinal waves and need a medium to travel.
Sound is incarnated on the medium by pressure, therefore, it is not surprising to believe that the medium and sound cannot be separated.
 One of the most interesting topics in acoustics for years is the
 extraordinary acoustic transmission (EAT) through subwavelength apertures in plates
 after a series of pioneering works in extraordinary optical transmission (EOT)
  \cite{bethe,genet,liu}.
 It has been known that the extraordinary transmission of sound
 is similar to its optical counterpart EOT \cite{zhang,lu}.
 Most of the previous works were the EAT of many holed plates or
a large transmission at a specific frequency \cite{hou,estrada1,estrada2,wang}.
However, an anti-transmission that allows a medium to pass through
a holed structure but blocks the sound has not been studied very much.

In recent years the research on EAT has accelerated
due to the fast development of acoustic metamaterials.
 The separation between sound and medium has been successfully demonstrated \cite{fang}.
The key is the Helmholtz resonators.
It makes the effective bulk modulus of the medium negative.
It is not a new system, but a new interpretation of the acoustic resonance.
If the medium and sound are separated,
we cannot hear the sound because the sound wave cannot travel without a medium,
but we can still feel the medium or air pressure because the medium can travel without the sound.
With the help of a modified Helmholtz resonator,
we designed a soundproof window or wall where air is allowed to flow freely
within some specific frequency ranges.
The air transparent window is a medium-sound separator
and the reverse form of the EAT
- an extraordinary acoustic anti-transmission (EAAT).

\begin{figure}
\resizebox{!}{0.2\textheight}{\includegraphics{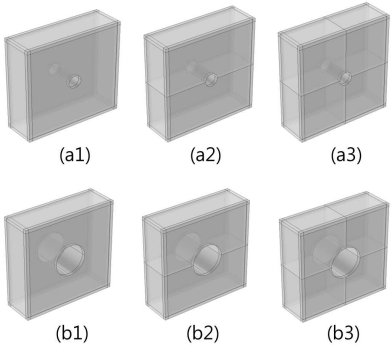}}
\caption{Artificial atoms of diffraction resonators.
Diameters of the air holes: 20$mm$ for (a1), (a2), and (a3),
 and 50$mm$ for (b1), (b2), and (b3).
There are three structures: one room for (a1) and (b1), two rooms for (a2) and (b2),
and four rooms for (a3) and (b3).
  }
 \label{atom}
\end{figure}

There are two conditions for a soundproof window where air passes through freely.
The first condition is strong diffraction.
It makes the wave diffuse into the resonator.
A common bottle-type Helmholtz resonator is not suitable for the purpose.
Therefore, we designed another resonator of an artificial atom
and called it a {\em diffraction resonator} in Fig.~\ref{atom}.
It has an air hole in the center of the body to maximize the diffraction effect.
The atom is composed of one, two, or four resonators.
  The air hole and the resonator are separated
 by an air filter \cite{filter} to match the acoustic impedance.
 Without the filtering it may create the noise of a wind instrument.
 Afterwards, we composed the artificial atoms in parallel and series to construct the window.
 It is made of transparent acrylic of a thickness of $5mm$.
   The inner dimension of each atom is $150mm \times 150mm \times 40mm.$
    The diameters of the air holes are $20mm$ and $50mm$.
It is different from a perforated sheet or the acoustic pipes of low-pass filters \cite{kut,yuya}.
 The $20mm$ window is designed to measure  background noise level of complete soundproofing
 while the $50mm$ window serves the practical purpose of noise reduction.

 The diameter (D) of the air hole of the artificial atoms should be much smaller than
the wavelength of the applied sound wave $\lambda$ for strong diffraction.
Therefore, the upper limit of the soundproof frequency is restricted
by the diffraction condition
\begin{equation}
f < f_D,
 \label{1}
\end{equation}
where $f_D$ is the cutoff frequency of the strong diffraction.

\begin{figure}
\resizebox{!}{0.18\textheight}{\includegraphics{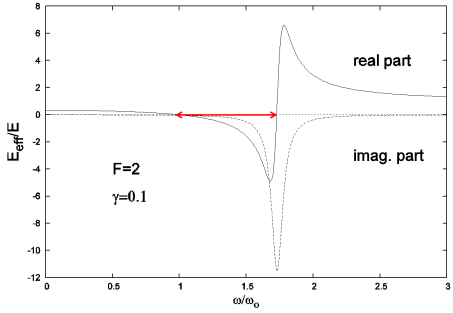}}
\caption{A typical plot of effective modulus around resonance.
The arrowed region of the frequency has a negative real part of the modulus.
 }
 \label{modulus}
\end{figure}

The second condition is the negative bulk modulus of the resonators that separate
the sound from its medium.
Bulk modulus $E$ is defined by $\Delta P=-E \Delta V/V$,
where $V$ is the volume.
The speed of an acoustic wave in fluid depends on the fluid's compressibility and inertia.
If the fluid has the bulk modulus $E$ and an equilibrium density $\rho$,
then the speed of sound in it is $v = \sqrt{E/\rho}$.
The resonance of accumulated waves in the  resonator
reacts against the applied pressure at some specific frequency ranges.
The negative modulus is then realized by passing the acoustic wave through the resonators.
By this procedure the amplitude of the sound wave is attenuated exponentially
by imaginary velocity or imaginary wave-vector.
    The general form of the effective bulk modulus of acoustic waves
 is given by the complex form similar with the effective electric permittivity \cite{fang,lee}
\begin{equation}
E_{eff}^{-1}={E}^{-1}\left[ 1-\frac{F \omega_o^2}{\omega^2 -
\omega_o^2 + i \Gamma \omega} \right],
 \label{bulk}
\end{equation}
where $\omega_o$ is the resonance frequency,  $\Gamma$ is a loss by damping,
 and $F$ is a geometric factor.

    We plotted the real and imaginary part of the effective modulus in Fig.~\ref{modulus}.
  The negative range of the real part is the stop-band of the wave.
The acoustic intensity decays at the frequency range.
When the loss is small, the effective bulk modulus has a negative real value at the frequency range
\begin{equation}
f_o < f < \sqrt{1+F} f_o,
\label{5}
\end{equation}
where $f_o = \omega_o/2\pi$.
This is the second condition for soundproofing.
    The soundproof range depends on the magnitude $F$.
The $F$ is the ratio of the volume of the resonator compared with the volume of the air passage,
and will be estimated by a combined method of theory and experiment.

\begin{figure}
\resizebox{!}{0.14\textheight}{\includegraphics{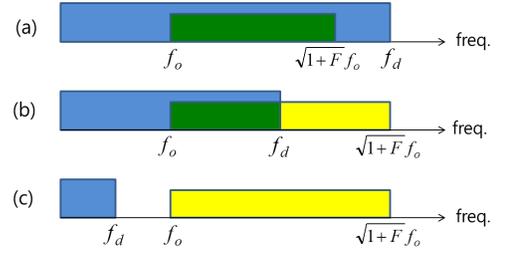}}
\caption{The blue one is the region of the diffraction
and the yellow one is the region of the negative modulus.
The overlapped green one is the soundproof region.
 }
 \label{range}
\end{figure}

We plotted the two conditions of the soundproof together in Fig.~\ref{range}.
For the soundproof window the two conditions of Eq.~(\ref{1}) and Eq.~(\ref{5}) must  both be satisfied.
 There is no soundproof in (c).
The smaller air hole and lower resonant frequency guarantee the wider frequency range of soundproofing.
  The larger air hole helps to pass through the air without losing the pressure
   but it lowers the cutoff frequency.

\begin{figure}
\resizebox{!}{0.12\textheight}{\includegraphics{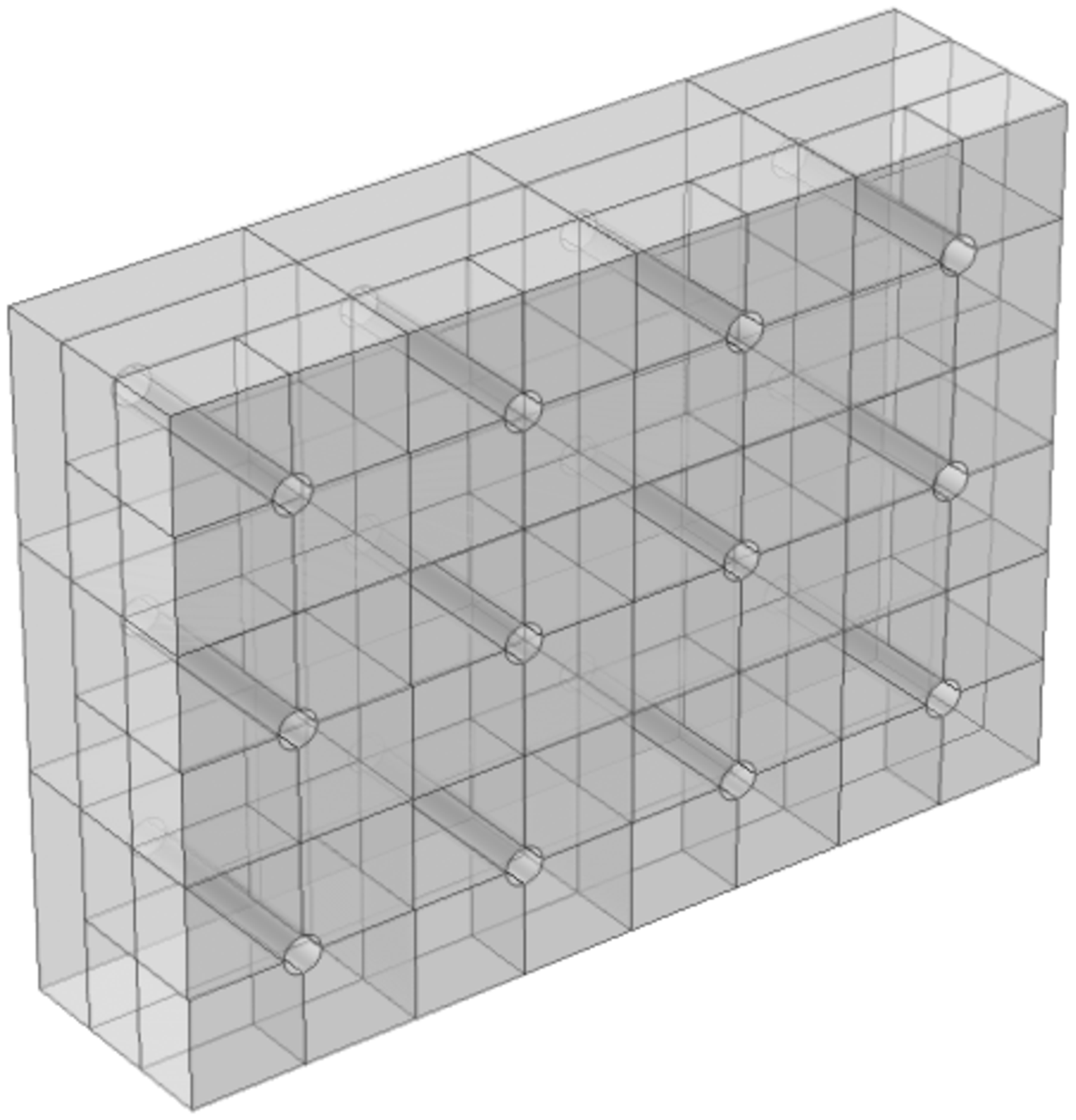}}
\resizebox{!}{0.14\textheight}{\includegraphics{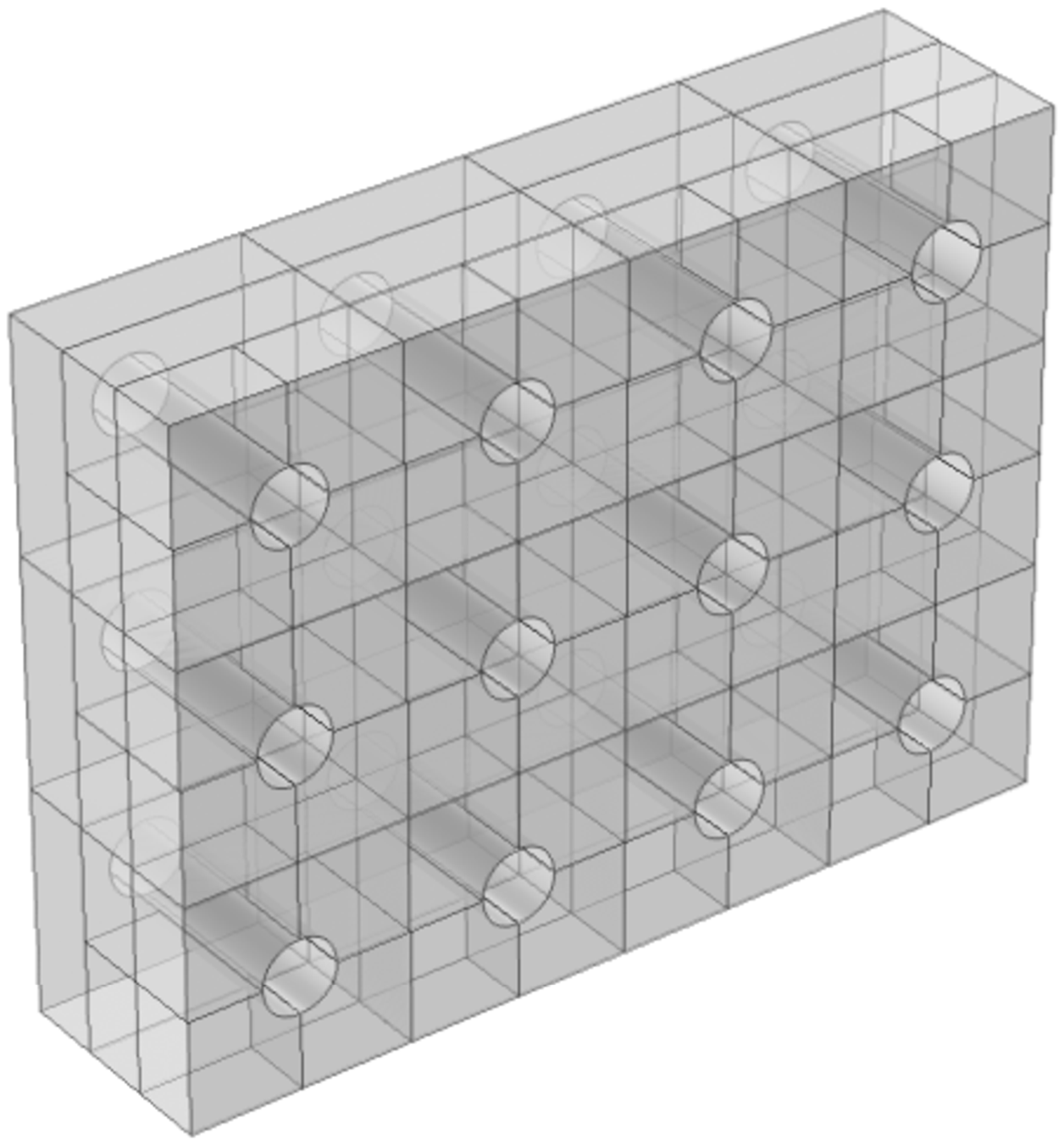}}
\caption{Designs of the medium-sound separator.  $20mm$ (left) and $50mm$ (right).
It is composed of the three kinds of artificial atoms which are connected in series and parallel.
 }
 \label{fig4}
\end{figure}

The sample soundproof window has an array of $3 \times 4 \times 3$ artificial atoms
arranged in parallel and series positions like the resonators in Fig.~\ref{fig4}.
  We connected three pieces of the artificial atoms in Fig.~\ref{atom} in series.
 The first piece has one room, the second one has two rooms, and the third one has four rooms.
 The number of rooms is the number of the resonators.
 The large volume or small entrance area corresponds to the low resonant frequency.
By this structure the window has three different stop-bands.

The minimum length of the air passage for the soundproof is obtained from the imaginary wave-vector.
The amplitude of the plane wave attenuates as
$e^{ikx} = e^{-2 \pi |n| x/\lambda},$
where $|n|$ is the refractive index in the resonator and is close to 1.
For the attenuation of the amplitude of $1/e$, the length of the air passage is $x=\lambda/2\pi$.
However, it is not necessary to be too long due to the background noise level.
The thickness or total pure length of the air passage of the windows is $40mm \times 3 =120mm$.
We can make the thickness of the window thinner if we curve the passage of the air.

\begin{figure}
\resizebox{!}{0.13\textheight}{\includegraphics{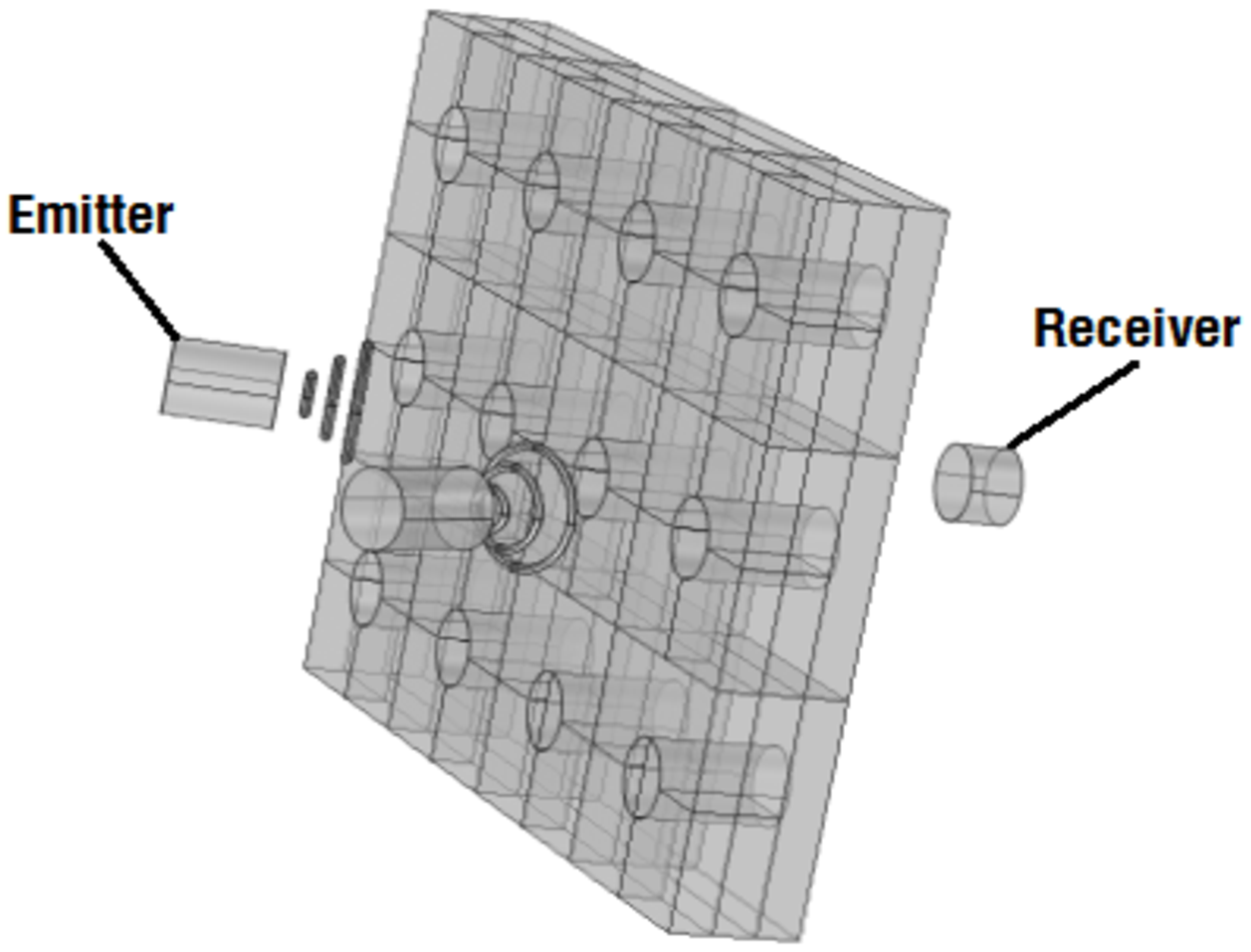}}
\resizebox{!}{0.10\textheight}{\includegraphics{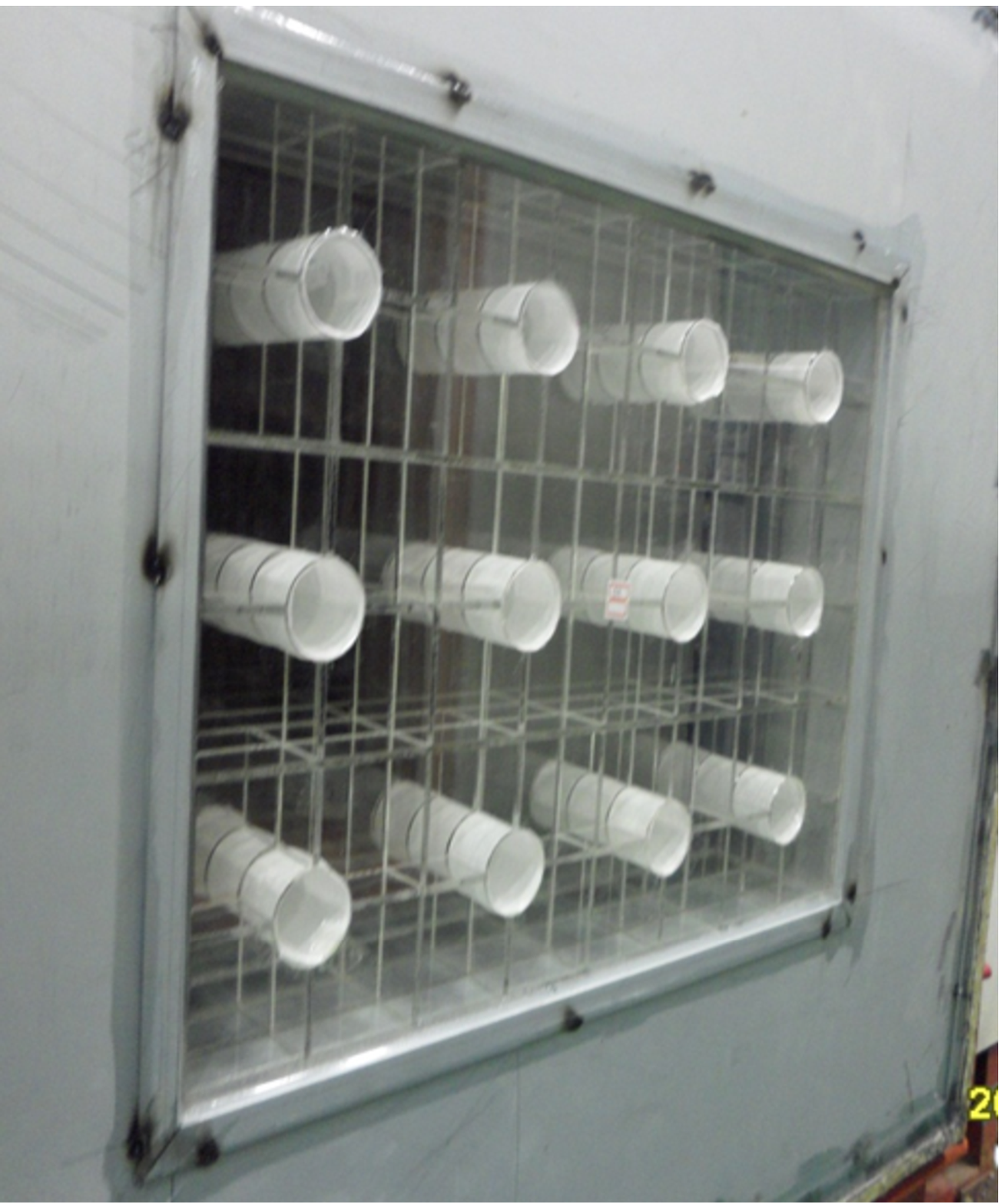}}
\caption{Diagram and picture of the measurement of the 50$mm$ window.
 }
 \label{fig5}
\end{figure}
\begin{figure}
\resizebox{!}{0.2\textheight}{\includegraphics{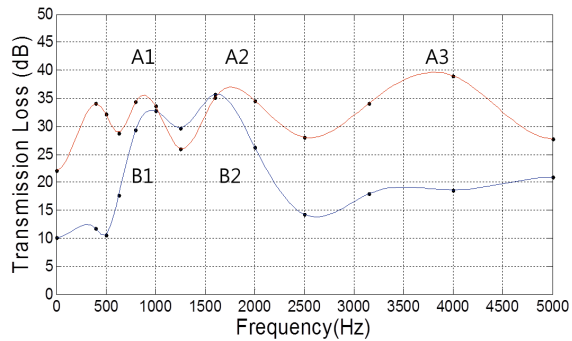}}
\caption{ Transmission loss of the air transparent soundproof windows.
The x-axis is 1/3 Oct. band center frequency.
The red line is for the $20mm$ window and the blue line is for the $50mm$ window.
 }
 \label{fig6}
\end{figure}
The measurement of sound transmission loss using small chambers has been carried out in accordance with the test standards for large reverberation rooms: ISO 10140:2010, ASTM E 90:2004. The test facility called as `mini-chamber' consists of two adjacent reverberant chambers with a test opening between them in which the test specimen is inserted:
(a)area of the specimen: $W 1.2m \times H 1.0m$,
(b)volume of Source Room: $2.808m^3$, and
(c)volume of Receiving Room: $3.252m^3$.
We applied the sound waves of $400 - 5,000Hz$ with a sound level of about $80dB$
from two emitters positioned diagonally with about $100$ degrees in Fig.~\ref{fig5}.
The distance between the emitters and the window is $1,200mm$
and the distance between the receiver and the window is $400mm$.

    The transmission losses for the two soundproof windows are plotted
in Fig.~\ref{fig6} with spline fits.
The small lowest frequency peak at around $300Hz$ may come from a periodic structure of the window.
The sound level is reduced by about $30 - 35$dB in the above frequency range with the $20mm$ window, and by about $20 - 35$dB in the frequency range of $700 - 2,200Hz$  with the $50mm$ window.
Each atom creates each peak in principle.
We connected three different atoms, but we can see only two peaks,
 B1 and B2 at the $50mm$ window.
The highest one of the $50mm$ window is cutoff by the diffraction condition in Eq.~(\ref{1}).
It means that the cutoff frequency is about $2,400Hz$ and it corresponds to
 $f_D \simeq v/3D$ .
The $20mm$ window corresponds to (a) and the $50mm$ window corresponds to (b) in Fig.~\ref{range}.

The resonance frequency of the artificial atoms is obtained approximately from
that of the bottle-type Helmholtz resonator \cite{bera,kut}.
\begin{equation}
f_o=\frac{v}{2\pi}\sqrt{\frac{S_H}{l_{e}V_H}}
 \simeq \frac{v}{2\pi}\sqrt{\frac{\pi r_e}{0.85 V}}
= \frac{v}{2\pi}\sqrt{\frac{\pi\sqrt{NDt} }{0.85 V}}.
\label{9}
\end{equation}
$S_H$ and $V_H$ are the neck area and volume of the Helmholtz resonator.
$S$ and $V$ are the surface area of the entrance and the volume inside of the artificial atom.
Then, we have $V_H=V/N$  and $S=\pi D t/N \equiv \pi r_{e}^2$,
where $D$ is the diameter of the air hole,
 $t$ is the thickness of the atom or the length of the air passage,
 and $N$ is the number of rooms or number of resonators.
 $l_{e}$ is the effective length of the neck given as $l_{e} = l + 0.85 r_{e}$
 and $r_{e}$ is the effective radius of the entrance of the diffraction resonator \cite{bera}.
 It gives the effective radius as $r_e = \sqrt{Dt/N}$.
Note that   $l_{e} = l + 0.85 r_{e} \simeq  0.85 r_{e}$ since $l << r_{e}$.

\begin{table}[h]
   \caption{Resonance frequencies and the corresponding geometric factors.}
     \label{tab:1}
  \centering
 \begin{tabular}{|c|c|c|c|c|}
    \hline \hline
    resonator &  rooms & $  f_o  (Hz)  $  &  $f_{peak}(Hz)$ & \hspace{0.15in} $F$ \hspace{0.2in} \\
    \hline \hline
    A1 & 1 &  590 &   900  & 3.3  \\
    \hline
    A2 & 2 & 700  & 1700  & 14 \\
    \hline
    A3 & 4 & 830  & 3800  & 62 \\
    \hline
     B1 & 1 & 770   & 950  & 1.2 \\
    \hline
   B2 & 2 & 910  & 1600  & 5.3 \\
    \hline \hline
   \end{tabular}
\end{table}

The geometric factor $F$ is estimated  from Eq.~(\ref{5})
if the stop-band in Fig~\ref{bulk} is symmetric
 \begin{equation}
\sqrt{1+F}=\frac{ 2f_{peak}}{f_o}-1,
 \label{geo}
\end{equation}
where $f_{peak}$ is the peak frequency and read from the Fig.~\ref{fig6}.
$f_o$ is obtained from Eq.~(\ref{9}).
We can estimate the geometric factor $F$  in Eq.~(\ref{bulk})
of every atom in Fig.~\ref{atom}.
We summarized the $F$ in Table 1.
If we use various air filters, we may have different results.
We observe that the geometric factor $F$ increases as the number of rooms increases.
$F$ is roughly proportional to the square of the number of rooms.


 We have presented an extraordinary acoustic anti-transmission (EAAT) through subwavelength apertures.
For air transparent soundproof, we suggested two conditions.
One is the strong diffraction and the other is the effective negative modulus.
Through applying the two conditions, we have designed a soundproof window where air passes through freely.
  The window is composed of many artificial atoms or diffraction resonators
  connected in series and parallel.
  The diameter of the hole in the artificial atoms should be much less than the wavelength
of the sound wave for a strong diffraction; the diameters of the holes being $20mm$ and $50mm$.
We created the resonator to intercept sound from the sound waves at some frequency ranges
which led to a separation of the medium and sound.
Afterwards, we applied sound waves in the range of $400 - 5,000Hz$ to the two windows.
We then observed a serious transmission loss of sound within specific frequency ranges.
The loss was $20 - 35$dB with the $50mm$ window in the range of  $700 - 2,200Hz$.

This air transparent window was not designed for complete soundproof but for noise reduction.
 The frequency range of soundproofing could be adjusted by several factors
 such as the volume of the resonator, the entrance area of the resonator,
   the size of the air hole, the property of the air filter, etc.
The thickness of the window could be very thin if we used a curved passage of air.
The geometric factor of the effective bulk modulus was estimated.
However, we needed to improve it by
considering the properties of the air filter and the asymmetry of the stop-band.

The structure of the air transparent soundproof window or wall is so simple that any carpenter can make it.
 The soundproof frequency range is tunable.
 There is a wide range of application areas
such as soundproof windows of houses close to noisy area,
the soundproof walls in residential areas, etc.
For example, if we are in a combined area of sounds from sea waves of low frequency
 and noises from machine operating at a high frequency,
 we can hear only the sounds from sea waves with fresh air.
 These principles should work in water as well as in air and may contribute to
underwater noise reduction for marine life.

We send many thanks to professors S. H. Lee and N. Fang for useful discussions.
This research was supported by Basic Science Research Program
through the National Research Foundation of Korea(NRF)
funded by the Ministry of Education, Science and Technology(2013-022016).

\end{document}